\documentclass[10pt,conference]{IEEEtran}

\RequirePackage{ifthen}
\newboolean{debugLinks}\setboolean{debugLinks}{false}
\newboolean{debugCites}\setboolean{debugCites}{false}
\newboolean{debugPages}\setboolean{debugPages}{false}

\pdfpageattr{/Group << /S /Transparency /I true /CS /DeviceRGB >>}

\usepackage[utf8]{inputenc}
\usepackage[T1]{fontenc}
\usepackage{textcomp}
\usepackage{microtype}
\usepackage{calc}
\usepackage[skins]{tcolorbox}

\usepackage[normalem]{ulem}
\usepackage{footmisc}

\usepackage{multirow}

\usepackage{lipsum}

\usepackage{silence}
\usepackage{csquotes}
\ifthenelse{\boolean{debugCites}}{
    \usepackage[backend=biber,style=verbose,bibstyle=ieee,doi=true,isbn=false,mincitenames=1,maxcitenames=2,minbibnames=1,maxbibnames=60]{biblatex}
}{
    \usepackage[backend=biber,style=ieee,bibstyle=ieee,doi=false,isbn=false,mincitenames=1,maxcitenames=2,minbibnames=1,maxbibnames=60]{biblatex}
}
\DeclareFieldFormat{sentencecase}{#1} 
\DeclareFieldFormat{titlecase}{#1} 
\usepackage{xpatch}

\xpatchbibmacro{textcite}{\addspace}{\addnbspace}{}{}
\xpatchbibmacro{Textcite}{\addspace}{\addnbspace}{}{}
\DeclareDelimFormat{namelabeldelim}{\addhighpenspace}

\DefineBibliographyStrings{english}{
    andothers = et~al\adddot\addspace
}

\DeclareSourcemap{
    \maps[datatype=bibtex]{
        \map[overwrite=true]{
            \step[fieldsource=school,fieldtarget=institution]
        }
    }
}
\RenewDocumentCommand\cite{O{}m}{%
    ~\autocite[#1]{#2}
}
\NewDocumentCommand\tablecite{O{}m}{%
    \autocite[#1]{#2}
}

\newcommand{\thisReferenceHasNoDoiAndThereIsNothingICanDoAboutIt}[1]{%
    \expandafter\newcommand\csname sommer@missingdoiwhitelist@#1@okay\endcsname{}%
}
\newcommand{\printmissingdoierror}[1]{%
    \ifcsname sommer@missingdoiwhitelist@#1@okay\endcsname%
    \else%
        \iffieldundef{doi}{%
            \todo{missing DOI for #1}%
            \BibliographyWarning{#1: missing DOI}%
        }{}%
    \fi%
}
\AtEveryCitekey{\printmissingdoierror{\thefield{entrykey}}}

\thisReferenceHasNoDoiAndThereIsNothingICanDoAboutIt{haykin2008communication}
\thisReferenceHasNoDoiAndThereIsNothingICanDoAboutIt{proakis1994communication}
\thisReferenceHasNoDoiAndThereIsNothingICanDoAboutIt{ziemer2014principles}
\thisReferenceHasNoDoiAndThereIsNothingICanDoAboutIt{rappaport2010wireless}
\thisReferenceHasNoDoiAndThereIsNothingICanDoAboutIt{proakis2001digital}
\thisReferenceHasNoDoiAndThereIsNothingICanDoAboutIt{almers2005solutionmanual}

\usepackage{amsmath}
\usepackage{upgreek}

\usepackage{amssymb}
\usepackage{amsfonts}
\usepackage{amsthm}

\usepackage[range-phrase=--,per-mode=symbol-or-fraction,range-units=single,list-units=single,detect-all,forbid-literal-units=true]{siunitx}
\AtBeginDocument{
\DeclareSIUnit\bit{bit}
\DeclareSIUnit\byte{Byte}
\DeclareSIUnit\decibeli{dBi}
\DeclareSIUnit\decibelm{dBm}
\DeclareSIUnit\kmh{\kilo\meter\per\hour}
\DeclareSIUnit\mbps{\mega\bit\per\second}
\DeclareSIUnit\mph{mph}
\DeclareSIUnit\mw{\milli\watt}
\DeclareSIUnit\resourceblock{RB}
\DeclareSIUnit\vehicle{veh}
\DeclareSIUnit\watthour{Wh}
\DeclareSIUnit\usdollar{US\$}
\DeclareSIUnit\pplong{\acf{PP}}
\DeclareSIUnit\ppshort{\acs{PP}}
}

\usepackage{booktabs}
\usepackage[]{url}

\makeatletter
\def\url@cmsstyle{%
  \def\UrlSpecials{%
    \do\-{\mathchar`-}%
  }%
}
\makeatother

\usepackage[inline]{enumitem}

\usepackage[caption=false,font=footnotesize]{subfig}
\usepackage{graphicx}
\DeclareGraphicsExtensions{.pdf,.png,.jpg,.jpeg}
\usepackage{xfrac}

\usepackage{tikz}
\usetikzlibrary{arrows}
\usetikzlibrary{calc}
\usetikzlibrary{chains}
\usetikzlibrary{scopes}

\ifthenelse{\boolean{debugCites} \OR \boolean{debugLinks} \OR \boolean{debugPages}}{
    \WarningsOff[everypage]
    \usepackage[placement=top]{background}
    \SetBgContents{draft}
    \SetBgScale{1}
    \SetBgVshift{-7}
}{
}

\usepackage[american]{babel}
\usepackage{hyphenat}

\usepackage{algorithmic}
\usepackage{algorithm}

\usepackage[capitalize,noabbrev,nameinlink]{cleveref}
\crefformat{footnote}{#2\textsuperscript{\normalfont #1}#3}
\crefformat{enumi}{#2#1#3}
\usepackage{footmisc}

\RequirePackage{xstring}
\RequirePackage{xparse}
\RequirePackage[nogroupskip,nonumberlist,nopostdot,acronyms,shortcuts,nomain]{glossaries-extra}
\RequirePackage{glossaries-prefix}
\setabbreviationstyle[acronym]{long-short}
\makenoidxglossaries
\glsdisablehyper  

\newglossarystyle{sommer}{%
  \setglossarystyle{list}%
}

\renewcommand{\glossarysection}[2][]{}

\NewDocumentCommand\acrodef{mmO{}}{%
    \newacronym[#3]{#1}{#1}{#2}%
    \glsfielddef{#1}{prefix}{a }
    \glsfielddef{#1}{prefixfirst}{a }%
}

\renewcommand{\acf}[1]{\glsentryfull{#1}}

\NewDocumentCommand\acroplural{mom}{%
  \glsfielddef{#1}{longpl}{#3}%
  \glsfielddef{#1}{firstpl}{#2}%
}

\NewDocumentCommand\acresetall{}{%
    \glsresetall%
}

\usepackage{todonotes}

\usepackage[noadjust,noparboxrestore]{marginnote}
\makeatletter
\def\todo{%
    \begingroup%
    \color{magenta}%
    \ifnum\@floatpenalty<0\relax%
    \else%
        \setlength{\columnsep}{2cm} 
        \marginnote{\color{magenta}\rule{2pt}{1em}}%
        \obeylines%
        \obeyspaces%
        \begingroup\lccode`~=`\^^M\lowercase{\endgroup\def~}{\par\leavevmode}%
        \parindent0em%
        \catcode`\_=\active%
        \catcode`\<=\active\lccode`~=`<\lowercase{\def~}{$<$}%
        \catcode`\>=\active\lccode`~=`>\lowercase{\def~}{$>$}%
        \catcode`\#=\active\lccode`~=`\#\lowercase{\def~}{$\#$}%
        \catcode`\^=\active\lccode`~=`\^\lowercase{\def~}{$\hat{~}$}%
        \catcode`\&=\active\lccode`~=`\&\lowercase{\def~}{\&}%
    \fi%
    \todoCtd%
}\def\todoCtd#1{%
    TODO: #1%
    \ifx&#1&...\fi%
    \ifnum\@floatpenalty<0\relax%
    \else%
    \fi%
    \endgroup%
    \relax%
}
\makeatother

\NewDocumentCommand\IEEE{ s m >{\SplitArgument{4}{/}}d[] }{%
    \IfBooleanTF{#1}{}{IEEE\,}
    \nolinebreak[2]
    #2%
    \IfNoValueTF{#3}{%
    }{%
        \sommerIEEELettersSlashed#3%
    }%
}
\newcommand{\sommerIEEELettersSlashed}[5]{%
    \IfNoValueTF{#2}{%
    }{%
        \nolinebreak[3]
    }%
    #1%
    \IfNoValueTF{#2}{}{/#2}%
    \IfNoValueTF{#3}{}{/#3}%
    \IfNoValueTF{#4}{}{/#4}%
    \IfNoValueTF{#5}{}{/#5}%
}

\NewDocumentCommand\sommerInvisibleText{ m }{%
    {
    \fontsize{.1}{.1}\selectfont
    \pdfliteral page{q 3 Tr}
    #1%
    \pdfliteral page{Q}
    }
}

\NewDocumentCommand\sommerMultiUrl{ d[] >{\SplitArgument{4}{,}}m d[] }{%
    \texttt{%
    \def\sommerMultiUrlFront{\IfNoValueTF{#1}{}{#1}}%
    \def\sommerMultiUrlEnd{\IfNoValueTF{#3}{}{#3}}%
    \def\sommerMultiUrlSpace{\;}%
    \sommerMultiUrlMiddle#2%
    }%
}
\newcommand{\sommerMultiUrlMiddle}[5]{%
    \sommerMultiUrlFront{}%
    \sommerMultiUrlSpace{}\{\sommerMultiUrlSpace{}%
    \IfNoValueTF{#1}{}{\sommerInvisibleText{\sommerMultiUrlFront{}}#1}%
    \IfNoValueTF{#2}{}{\sommerInvisibleText{\sommerMultiUrlEnd{}}\sommerMultiUrlSpace{},\sommerMultiUrlSpace{}\sommerInvisibleText{\sommerMultiUrlFront{}}#2}%
    \IfNoValueTF{#3}{}{\sommerInvisibleText{\sommerMultiUrlEnd{}}\sommerMultiUrlSpace{},\sommerMultiUrlSpace{}\sommerInvisibleText{\sommerMultiUrlFront{}}#3}%
    \IfNoValueTF{#4}{}{\sommerInvisibleText{\sommerMultiUrlEnd{}}\sommerMultiUrlSpace{},\sommerMultiUrlSpace{}\sommerInvisibleText{\sommerMultiUrlFront{}}#4}%
    \IfNoValueTF{#5}{}{\sommerInvisibleText{\sommerMultiUrlEnd{}}\sommerMultiUrlSpace{},\sommerMultiUrlSpace{}\sommerInvisibleText{\sommerMultiUrlFront{}}#5}%
    \sommerInvisibleText{\sommerMultiUrlEnd{}}%
    \sommerMultiUrlSpace{}\}\sommerMultiUrlSpace{}%
    \sommerMultiUrlEnd{}%
}

\newcommand{\defineComputedValue}[2]{%
  \expandafter\newcommand\csname sommer-cv-#1\endcsname{#2}%
}
\newcommand{\computedValue}[1]{%
    \ifcsname sommer-cv-#1\endcsname%
        \csname sommer-cv-#1\endcsname%
    \else%
        \errmessage{Computed value `#1' not defined.}%
    \fi%
}

\usepackage{listings}

\acrodef{AI}{Artificial Intelligence}
\acrodef{CoT}{Chain-of-Thought}
\acrodef{LLM}{Large Language Model}
\acrodef{NLP}{Natural Language Processing}
\acrodef{GT}{Ground Truth}
\acrodef{QA}{Question-Answer}
\acrodef{eCDF}{Empirical Cumulative Distribution Function}
\acrodef{RAG}{Retrieval-Augmented Generation}

\usepackage{fancyhdr}
\fancypagestyle{firstpage}{%
  \fancyhf{} 

  \fancyfoot[L]{\scriptsize
    \copyright{} 2025 IEEE. This is the author's version of the paper accepted at IEEE Globecom Workshops 2025. 
    The final version is published in IEEE Xplore: \url{https://doi.org/10.1109/GCWkshps.2025.xxxxx}
  }
}

\csname endofdump\endcsname

\begin{document}

\setlength{\columnsep}{0.23in}

\title{Evaluating Open-Source Large Language Models for Technical Telecom Question Answering}

\author{%
    \IEEEauthorblockN{%
        Arina Caraus\IEEEauthorrefmark{1}\IEEEauthorrefmark{2}%
        ,
        Alessio Buscemi\IEEEauthorrefmark{1}%
        ,
        Sumit Kumar\IEEEauthorrefmark{1}%
        , and
        Ion Turcanu\IEEEauthorrefmark{1}%
    }%
    \IEEEauthorblockA{
        \IEEEauthorrefmark{1}Luxembourg Institute of Science and Technology (LIST), Luxembourg%
        \\
        \IEEEauthorrefmark{2}RMT Labs, Luxembourg%
    }%
    \small{
        \texttt{
            acaraus@rmt-labs.com
        }
        \texttt{
            \{~%
                alessio.buscemi%
                ,
                sumit.kumar%
                ,
                ion.turcanu%
            ~\}@list.lu
        }
    }
}


\maketitle%

\thispagestyle{firstpage}

\ifthenelse{\boolean{debugPages}}{
\thispagestyle{plain}
\pagestyle{plain}
}

\begin{abstract}\nohyphens{%
\acp{LLM} have shown remarkable capabilities across various fields.
However, their performance in technical domains such as telecommunications remains underexplored.
This paper evaluates two open-source \acp{LLM}, Gemma 3 27B and DeepSeek R1 32B, on factual and reasoning-based questions derived from advanced wireless communications material.
We construct a benchmark of 105 question–answer pairs and assess performance using lexical metrics, semantic similarity, and LLM-as-a-judge scoring.
We also analyze consistency, judgment reliability, and hallucination through source attribution and score variance.
Results show that Gemma excels in semantic fidelity and \ac{LLM}-rated correctness, while DeepSeek demonstrates slightly higher lexical consistency.
Additional findings highlight current limitations in telecom applications and the need for domain-adapted models to support trustworthy \ac{AI} assistants in engineering.
}\end{abstract}

\acresetall%

%
\acresetall%

\section{Introduction}\label{sec:intro}


\acp{LLM} are sophisticated \ac{AI} models whose popularity is rapidly increasing, with successful applications already demonstrated in fields such as psychology\cite{hu2024psycollm}, medicine\cite{xie2024me}, law\cite{siino2025exploring}, and finance\cite{wu2023bloomberggpt}.
Despite this widespread adoption, their use in the telecommunications sector remains relatively limited.
Nonetheless, recent efforts have emerged to develop domain-specific \acp{LLM} aimed at enhancing tasks such as customer service automation, dynamic network configuration, and traffic classification.
The limited integration of \acp{LLM} in telecommunications can be attributed to the domain’s inherent complexity.
Unlike applications that rely primarily on linguistic and contextual reasoning, telecom tasks demand a solid understanding of physical and mathematical principles, such as electromagnetic wave propagation and modulation schemes.
Effective deployment in this space requires models to interpret formal systems, follow protocol specifications, and reason over signal processing operations.
These demands stretch the capabilities of current general-purpose models and underscore the need for specialized solutions tailored to the domain’s unique requirements.

Although recent \ac{LLM} families~-- such as GPT, Llama, DeepSeek, Gemma, and Mistral~-- have shown rapid progress, their ability to process highly technical, domain-specific content in telecommunications remains largely untested.
In particular, it is unclear whether such models can produce accurate and meaningful responses to questions requiring deterministic reasoning, structured derivations, and textbook-based knowledge.

In this paper, we aim to fill this gap by constructing and using a rigorous evaluation benchmark derived from the textbook \textit{Wireless Communications} by \textcite{molisch2011wireless}, a foundational reference in both academic and industrial settings.
Verified answers are sourced from the textbook’s official solution manual to ensure high fidelity and precision.
The resulting dataset provides a robust testbed for measuring \ac{LLM} performance on factual \ac{QA} in the telecommunications domain.
The goal of this work is to assess whether \acp{LLM} can act as reliable assistants for telecom engineers and operators.

In addition, trust is a fundamental prerequisite. Knowing that a model can produce correct answers in principle is not sufficient: if its outputs are inconsistent across repetitions of the same question, the user may unknowingly rely on flawed information, potentially leading to incorrect technical decisions.
For this reason, we place a strong emphasis not only on the \textit{correctness} of the answers, but also on the \textit{consistency} with which models provide them.
This consistency is measured both in answer generation and in the evaluation process when models serve as judges of quality.

Our study is guided by the following research questions:
\begin{itemize}
\item \textbf{RQ1:} Can \acp{LLM} provide technically accurate and complete answers to telecommunications questions?
\item \textbf{RQ2:} What metrics best capture the quality and utility of the answers generated by \acp{LLM}?
\item \textbf{RQ3:} How consistent are open-source \acp{LLM} in both generating answers and evaluating answer quality when exposed to the same technical questions in the telecommunications domain?
\end{itemize}

To address these questions, we develop a structured evaluation framework.
For RQ1, we assess whether \acp{LLM} can produce technically accurate and complete responses to questions covering diverse areas of telecommunications.
For RQ2, we evaluate answer quality using a combination of metrics: lexical similarity, semantic similarity, and assessments provided by a separate \ac{LLM} acting as a judge.
RQ3 is addressed by analyzing how consistent the models are in their generated responses and in the evaluations they provide when presented with identical technical prompts.
This study provides one of the first comprehensive assessments of state-of-the-art \acp{LLM} on technical telecommunications content across multiple areas.


%

\section{Related Work}\label{sec:relwork}

Recent research has explored the capabilities and limitations of \acp{LLM} in both general-purpose and domain-specific contexts. 
\Textcite{zhou2024large,maatouk2024large} point out that prompting techniques such as In-Context Learning (ICL) and Chain-of-Thought (CoT) often fall short in evaluate complex tasks such as multi-hop reasoning and factual consistency. 
General benchmarks like HellaSwag, SuperGLUE, and MMLU assess overall NLP performance but lack the granularity needed for expert-level knowledge in specialized fields like telecommunications.

In the telecom domain, \textcite{maatouk2025teleqna} introduced the TeleQnA dataset, evaluating models such as GPT-3.5 and GPT-4 with multiple-choice and open-ended questions from 3GPP standards and telecom-specific lexicons.
\Textcite{soman2023observations} focus on classifying technical documents from 3GPP's TSGs (RAN, SA, and CT), underscoring the deep technical expertise required for accurate categorization. 
However, most of these efforts concentrate heavily on 3GPP materials and use primarily multiple-choice formats, limiting their breadth\cite{bornea2024telco}.


\Textcite{bariah2024large} emphasize the growing importance of explainability in telecom-specific \acp{LLM}, especially as these models are integrated into network operations and decision-making systems.
Despite the utility of datasets like TeleQnA\cite{maatouk2025teleqna}, recent literature suggests that existing benchmarks do not fully capture the diversity of telecom knowledge\cite{zhou2024large}. 
\Textcite{ahmed2024linguistic}, for example, employed this benchmark to evaluate the performance of GPT-3.5 models on various telecom tasks.

Another emerging area of research is the development of conversational assistants tailored to telecom applications\cite{bariah2023understanding}.
These systems demand not only accuracy but also contextual understanding and dialog-level reasoning.

Our work expands upon prior efforts in three key ways: (1) we evaluate the lesser-studied DeepSeek model and compare its performance with Gemma; (2) we shift the focus from multiple-choice to free-form question answering; and (3) our dataset encompasses a broader spectrum of telecom topics beyond 3GPP, offering a more diverse and realistic benchmark for assessing \ac{LLM} performance in this domain.

\section{Methodology}\label{sec:method}
\label{sec:methodology}


\subsection{Dataset Construction}
\label{sub:dataset_construction}

We built a structured \ac{QA} dataset based on the textbook \enquote{Wireless Communications} (2nd ed.) by \textcite{molisch2011wireless}, chosen for its technical depth and domain relevance.
The dataset includes 105 questions, with $\approx \SI{60}{\percent}$ being conceptual and $\approx \SI{40}{\percent}$ requiring mathematical derivations or calculations.
Reference answers were extracted from the official Solution Manual for \enquote{Exercises in the Textbook Wireless Communication by A.F. Molisch}\cite{almers2005solutionmanual}, providing high-quality ground truth annotations.

\begin{figure}[b]
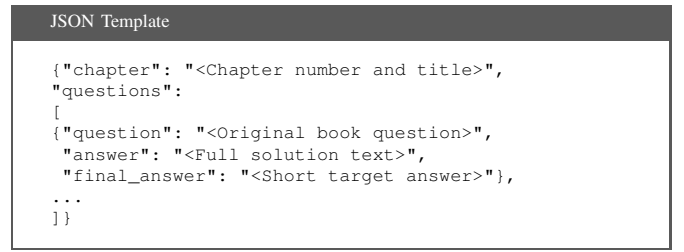

{\scriptsize
\begin{tcolorbox}[colback=white, colframe=black!65, title=JSON Template, sharp corners, boxrule=0.8pt]
\begin{verbatim}
{"chapter": "<Chapter number and title>",
"questions": 
[
{"question": "<Original book question>",
 "answer": "<Full solution text>",
 "final_answer": "<Short target answer>"},
...
]}
\end{verbatim}
\end{tcolorbox}}
\caption{Test Dataset JSON Template. This format defines the structure used to evaluate an LLM's performance. It includes a set of chapter-based questions, full solution texts, and expected short-form answers (used as ground truth) to compare against generated outputs.}
\label{box:solution-json}
\end{figure}

Each data instance is formatted as a JSON object organized by chapter, and includes three fields: \texttt{question} (the original exercise from the textbook), \texttt{answer} (the corresponding detailed solution from the manual), and \texttt{final\_answer} (a concise, manually curated summary that highlights key facts or numerical values).
The \texttt{final\_answer} field was specifically introduced to support evaluation tasks and to better align with the output structure of \acp{LLM}.

To ensure consistency, we retained only \ac{QA} pairs with both the \texttt{answer} and \texttt{final\_answer} in the solutions manual.
This filtering yielded a refined dataset of 105 high-quality \ac{QA} pairs spanning multiple chapters.
The resulting dataset provides a domain-specific benchmark for assessing the factual performance of \acp{LLM} in the context of wireless communications.
An example schema is shown in \cref{box:solution-json}.

\subsection{Evaluation Framework}
\label{sub:evaluation}

\begin{figure}
    \centering
    \includegraphics[width=\columnwidth]{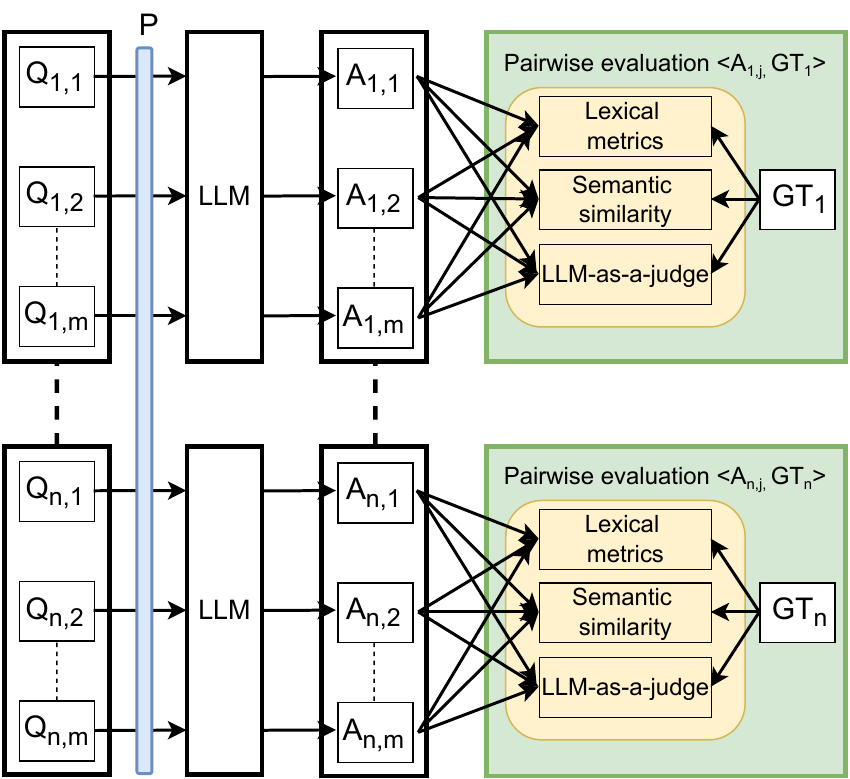}
    \caption{Evaluation pipeline: each question \( Q_i \) is submitted \( m \) times to the LLM, yielding answers \( A_{i,1}, \ldots, A_{i,m} \), each evaluated against the ground truth \( GT_i \) using lexical, semantic, and judgment-based metrics.}
    \label{fig:Evaluation per answer}
\end{figure}

To systematically assess the performance of \acp{LLM} on domain-specific technical questions, we implemented a multi-tiered evaluation framework that integrates both automatic metrics and model-based judgments.
As illustrated in \cref{fig:Evaluation per answer}, the framework operates at the level of individual \ac{QA} pairs and applies three complementary evaluation strategies: lexical similarity, semantic similarity, and \ac{LLM}-as-a-Judge scoring.

Each question \( Q_i \) in the dataset is submitted \( m \) times to an \ac{LLM} \( L \) using a structured prompt \( P \), described in \cref{subsec:prompting}.
In this work, we selected \( n = 105 \) questions, as described in \cref{sub:dataset_construction}, and set \( m = 20 \) repetitions per question.
This configuration was chosen to balance computational cost and statistical robustness, and to enable a consistency analysis of the model’s behavior across repeated queries.
The resulting answers \( A_{i,1}, \ldots, A_{i,m} \) are compared with the reference answer~-- i.e., \ac{GT}~-- through pairwise evaluation.

Lexical similarity is assessed using BLEU, ROUGE-L, and METEOR metrics, which capture surface-level token overlap but may penalize legitimate variation in expression.
Semantic similarity is computed via cosine distance between sentence embeddings of \( A_{i,j} \) and \( GT_i \), offering a meaning-oriented perspective on answer quality.
Finally, in the \emph{LLM-as-a-Judge} setup, a separate language model is prompted to evaluate each answer by directly comparing it with the ground truth \( GT_i \).
It assigns a score from 0 to 10, where 0 indicates a response that is entirely incorrect or irrelevant, and 10 corresponds to a perfectly aligned and factually accurate answer.

In addition to answer quality, we examine the reliability of \acp{LLM} through two forms of consistency.
Answer consistency captures the variability in responses across the \( m \) independent generations of the same question, offering insight into the model’s determinism.
Scoring consistency reflects the variation in repeated judgment-based evaluations of identical answers.
Together, these metrics provide a robust and interpretable assessment of both performance and behavioral stability~-- key requirements for telecom-grade applications.

\subsection{Prompting}
\label{subsec:prompting}

In this work, we explore two complementary prompting setups, \emph{zero-shot} and \emph{few-shot}, to examine the responses generated by the \ac{LLM}.
In the zero-shot configuration, the model is presented solely with the raw textbook question \(Q_i\) and is expected to produce an answer without any additional guidance or contextual examples.
An example of such an input is shown in \cref{fig:zero-shot}.

\begin{figure}[b]
\footnotesize
\begin{tcolorbox}[
    enhanced,
    width=0.48\textwidth,
    colframe=black,
    colback=white,
    boxrule=0.8pt,
    sharp corners,
    colbacktitle=black!65,
    coltitle=white,
    title={Zero-shot Input},
    left=0pt,right=0pt,top=2pt,bottom=2pt
]
\lstset{
  basicstyle       = \ttfamily\footnotesize, 
  frame            = none,
  breaklines       = true,
  breakatwhitespace= true,
  breakindent      = 0pt,
  columns          = fullflexible,
  xleftmargin      = 0pt
}
\begin{lstlisting}[%
  language=TeX,
  label={lst:zero-prompt-agent}
]
Assuming that directions of arrival are uniformly distributed at the MS, how large is the correlation coefficient (for a GSM-1800 system) between the channel in the middle and the end of the burst when the MS moves at 250~km/h? How large is the correlation coefficient between the channels at the beginning and end of the burst?
\end{lstlisting}
\end{tcolorbox}
\caption{An example of zero-shot input}
\label{fig:zero-shot}
\end{figure}

The few-shot setup follows a different approach.
Here, the same question \(Q_i\) is embedded in a structured instruction template, shown in \cref{fig:few-shot}, which includes two worked examples from the telecommunications domain to provide context and guidance.
The prompt also explicitly requires the model to generate only a concise \texttt{final\_answer}, and mandates citation of the source material.
To enhance response diversity and enable a more robust evaluation, we query the model $m=20$ times using this few-shot prompt, resulting in a response set \(\{R^{\text{LLM}}_{i,1},\dots,R^{\text{LLM}}_{i,m}\}\).
This ensemble of answers serves as the foundation for the pairwise and ground truth similarity analysis described in \cref{sec:method}.

\begin{figure}[b]
    \centering
    \footnotesize
    \begin{tcolorbox}[
        enhanced,
        width=0.48\textwidth,
        colframe=black,
        colback=white,
        boxrule=0.8pt,
        sharp corners,
        colbacktitle=black!65,
        coltitle=white,
        title={Few-shot Prompt Template},
        left=0pt,right=0pt,top=2pt,bottom=2pt
    ]
    \lstset{
    basicstyle       = \ttfamily\footnotesize, 
    frame            = none,
    breaklines       = true,
    breakatwhitespace= true,
    breakindent      = 0pt,
    columns          = fullflexible,
    xleftmargin      = 0pt
    }
    \begin{lstlisting}[%
    language=TeX,
    label={lst:prompt-agent}
    ]
    You are given a question related to telecommunications, networking, or signal transmission.
    Your task is:
    1. Analyse the question carefully.
    2. Provide only the final_answer with no additional explanation or reasoning.
    3. The final_answer must be concise, accurate, and directly  respond to the question.
    4. At the end give the source from where you obtained the information. Use the format- Source: 

    Examples
    --------
    Input Question: Which of the following systems cannot transmit in both directions (duplex or semiduplex): (i) cellphone, (ii) cordless phone, (iii) pager, (iv) trunking radio, (v) TV broadcast system?
    Output: pager, TV broadcast

    Input Question: Communication is to take place from one side of a building to the other as depicted in Figure 30.1, using 2-m-tall antennas. 
    Convert the building into a series of semi-infinite screens and determine the field strength at the receive antenna caused by diffraction using Bullington's method for  (a) f = 900 MHz, (b) f = 1 800 MHz, and (c) f = 2.4 GHz.
    Output: (a) 0.0155, (b) 0.0109, (c) 0.0095

    Input Question: {question_text}
    \end{lstlisting}
    \end{tcolorbox}
    \vspace{-0.2cm}
\caption{An example of a few-shot prompt template}
\label{fig:few-shot}
\end{figure}

\subsection{Additional details}

\begin{figure}[b]
    \centering
    \footnotesize
    \begin{tcolorbox}[
        enhanced,
        width=0.48\textwidth,
        colframe=black,
        colback=white,
        boxrule=0.8pt,
        sharp corners,
        colbacktitle=black!65,
        coltitle=white,
        title={Evaluation JSON Template},
        left=0pt,right=0pt,top=2pt,bottom=2pt
    ]
    \begin{verbatim}
    {
    "question": "<QA input>",
    "generated_answer": "<LLM reply>",
    "rating": "<Judge LLM comment>",
    "rating_value": <Judge score>,
    "source": "<LLM source or 'NaN'>",
    }
    \end{verbatim}
    \end{tcolorbox}
\caption{An example of an evaluation JSON template}
\label{box:eval-json}
\end{figure}

To assess the transparency and potential hallucination behavior of the \ac{LLM} when answering technical questions, we incorporate a mandatory source attribution component into the prompt.
Each prompt instructs the \ac{LLM} to append a source for the answer it provides.

Regarding the \ac{LLM}-as-a-Judge evaluation, we apply a rule-based outlier filtering mechanism focused solely on enforcing rating boundaries to ensure the validity of the evaluation scores.
Let $\mathcal{S} = \{ {s_{i,j}} \}$ ($i \in [1,n]$, $j \in [1,m]$) denote the set of scores assigned by the \ac{LLM}-as-Judge to the generated answers.
Each score $s_i$ must fall within the interval $[0, 10]$, as specified in the scoring instructions.
All scores $s_i$ that violate this constraint are discarded from further analysis.
This step is essential to ensure that metrics and aggregations reflect only interpretable and rule-compliant scores.
For the evaluation with \ac{LLM}-as-a-Judge, the data must be in the template shown in \cref{box:eval-json}.

\section{Performance Evaluation}\label{sec:simres}
\label{sec:performance_evaluation}


\subsection{Experimental setup}
\label{subsec:setup}

In this work, we use two advanced language models with complementary reasoning capabilities.
DeepSeek-R1 32B is a reasoning-focused model trained to generate explicit, multi-step rationales, making it suitable for tasks requiring logical transparency. Gemma 3 27B, by contrast, is a lightweight instruction-tuned model optimized for efficient, coherent responses in general-purpose applications.
Both models were deployed locally on a single Nvidia Tesla V100 (\SI{32}{\giga\byte}) GPU using the Ollama library in Python, and were pulled from Hugging Face.
All processing and evaluation was conducted using Python 3.11.

Before conducting the performance evaluation, we first verified whether the models had prior exposure to the textbook solutions used in our test set.
This step is essential to determine whether the models generate answers by reasoning through the problem or simply by recalling content memorized during training.
To this end, we selected a subset of questions and prompted each model multiple times, explicitly asking for the sources of their answers, as described in \cref{subsec:prompting}.

For DeepSeek-R1, none of the answers cited the textbook or its solutions manual.
Instead, references were consistently made to other general telecommunications literature\cite{haykin2008communication, proakis1994communication, ziemer2014principles}.
This indicates that DeepSeek-R1 was likely not exposed to the ground-truth answers during training.
By contrast, Gemma~3~27B occasionally cited the textbook from which the questions were drawn, but never mentioned the corresponding solutions manual.
However, it frequently referenced other works\cite{proakis2001digital,rappaport2010wireless}.
From this, we conclude that while Gemma may have encountered the source of the questions, neither model had access to the actual answers.
This gives us confidence that their performance reflects genuine reasoning ability rather than memorization.





\subsection{Accuracy analysis}
\label{subsec:accuracy}

The accuracy of answers generated by the \ac{LLM} is calculated taking into consideration the correctness of reasoning to find the right answer.
We analyze whether the reasoning followed by the \ac{LLM} aligns with the \ac{GT}.
To complement this analysis, we evaluate the similarity between the generated and reference answers at both lexical and semantic levels.

\subsubsection{Lexical}
We used normalized BLEU-4, ROUGE-L, and METEOR metrics for lexical similarity.
These metrics measure word and phrase overlap between candidate and \ac{GT} texts.
They focus on precision, recall, and matching n-grams, rewarding outputs that closely follow the reference structure.

\begin{figure}
    \centering
    \subfloat[DeepSeek]{\includegraphics[width=.9\columnwidth]{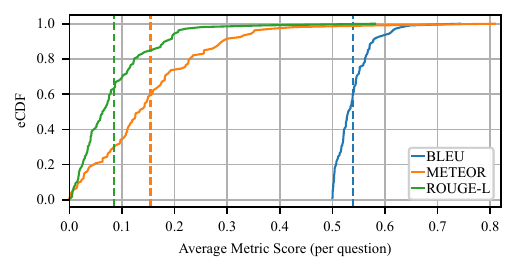}\label{subfig:deep-acc}}%
    \\
    \subfloat[Gemma]{\includegraphics[width=.9\columnwidth]{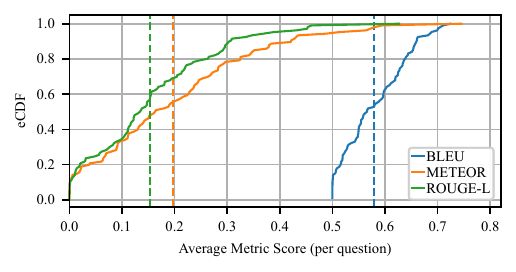}\label{subfig:gemma-acc}}%
    \caption{Accuracy analysis of answers $A_i$ generated by \ac{LLM} with Gemma and DeepSeek with respect to \ac{GT} for questions $Q_i$.}     
    \label{fig:accuracy}
    \vspace{-0.4cm}
\end{figure}
\Cref{fig:accuracy} shows that Gemma consistently achieves higher scores than DeepSeek across all three metrics.
Its score distributions are shifted toward the right, indicating a greater proportion of high-quality outputs.
For example, around \SI{40}{\percent} of Gemma's BLEU scores exceed 0.6, compared to only \SI{5}{\percent} for DeepSeek.
Similarly, more than \SI{20}{\percent} of Gemma's METEOR scores surpass 0.3, while only \SI{10}{\percent} of DeepSeek's scores surpass this threshold.
These results indicate that Gemma produces more precise, fluent, and structurally aligned responses, outperforming DeepSeek in both lexical and semantic terms.

\subsubsection{Semantic}
\begin{figure}[b]
    \centering
    \includegraphics[width=.9\columnwidth]{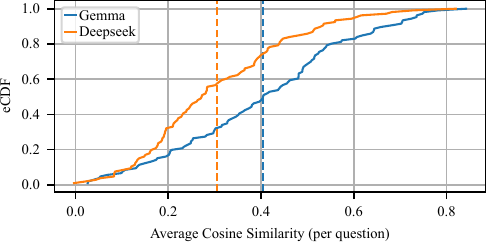}
    \caption{Accuracy analysis of answers  $A_i$ generated by \ac{LLM} with Gemma and DeepSeek using cosine similarity.}       
    \label{fig:accuracy_semantic}
\end{figure}
For semantic similarity, we use cosine similarity computed over sentence embeddings to assess whether the meaning conveyed by the generated answer is close to that of the \ac{GT}, even if the wording differs. 
Specifically, we use the all-MiniLM-L6-v2 model to produce the embeddings. 
This allows us to evaluate the deeper alignment in content beyond surface-level token matching.
The results illustrated in \cref{fig:accuracy_semantic} show that Gemma provides better performance in terms of semantic similarity, achieving an average cosine similarity of 0.4, compared to 0.3 achieved by DeepSeek.

\subsubsection{LLM-as-a-Judge}

\begin{figure}[b]
    \centering
    \includegraphics[width=.9\columnwidth]{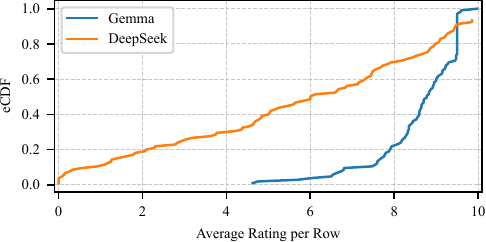}
    \caption{Accuracy analysis of answers $A_i$ generated by \ac{LLM} with Gemma and DeepSeek using LLM-as-a-Judge for scoring.}
    \label{fig:accuracy_llm}
\end{figure}
To complement the automatic evaluation, we also considered \ac{LLM}-as-a-Judge scores, where the same \ac{LLM} is used to both answer the question and evaluate the accuracy of the response.
The results, shown in \cref{fig:accuracy_llm}, indicate that Gemma again outperforms DeepSeek, with a greater proportion of its responses receiving higher ratings.
This subjective evaluation aligns well with the automatic metrics, confirming that Gemma's outputs are not only more lexically and structurally aligned with references, but also judged as more accurate and relevant by a powerful language model.

\subsection{Consistency analysis}
\label{subsec:consistency}

To assess the consistency of the \ac{LLM} when answering the same question multiple times, we repeated each question $Q_i$ exactly $m=20$ times.
We then computed the semantic and lexical similarity between each pair of consecutive responses, $\langle A_{i,j}, A_{i,j+1} \rangle$ for $j \in [1,m-1]$.
In addition, we evaluated variability by calculating the standard deviation of all 20 responses $A_{i,j}$ for each $Q_i$.
For the \ac{LLM}-as-a-Judge, we similarly measured consistency by computing the standard deviation of the scores it assigned across multiple judgments of the same answer.

\subsubsection{Lexical}

\begin{figure}[b]
    \centering
    \subfloat[Deepseek]{\includegraphics[width=0.9\columnwidth]{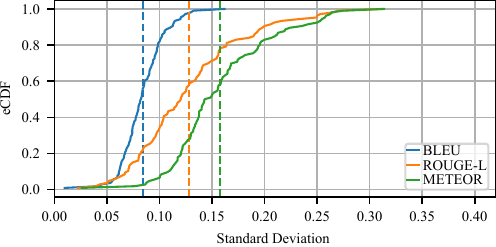}\label{subfig:deep-consistency}}%
    \\
    \subfloat[Gemma]{\includegraphics[width=0.9\columnwidth]{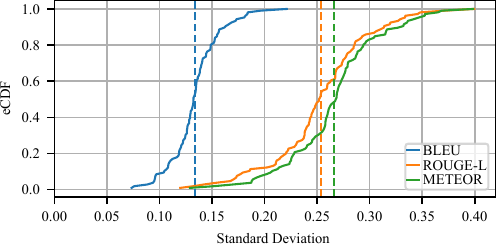}\label{subfig:gemma-consistency}}%
    \caption{Lexical consistency analysis of answers generated by \ac{LLM} with Gemma and DeepSeek calculated on pairs of answers $\langle A_i^j, A_i^{j+1} \rangle \forall Q_i$.}        
    \label{fig:consistency-lexical}
\end{figure}

To assess lexical and structural consistency, we measured the standard deviation of BLEU, METEOR, and ROUGE-L scores between consecutive outputs generated in response to the same prompt.
\Cref{fig:consistency-lexical} reveals that DeepSeek is consistently more stable than Gemma across all three metrics.
For example, over \SI{70}{\percent} of DeepSeek's BLEU pairwise deviations fall below 0.1, whereas Gemma's BLEU standard deviations are more broadly distributed, with many values exceeding 0.15.
The same pattern holds for ROUGE-L and METEOR, where DeepSeek’s curve rises steeply, indicating that a greater proportion of its outputs vary less between repeated runs.
These findings suggest that while Gemma may produce higher average-quality responses, DeepSeek delivers more reproducible outputs~-- a valuable trait in applications where deterministic behavior and response stability are critical.

\subsubsection{Semantic}
\begin{figure}[b]
    \centering
    \includegraphics[width=.9\columnwidth]{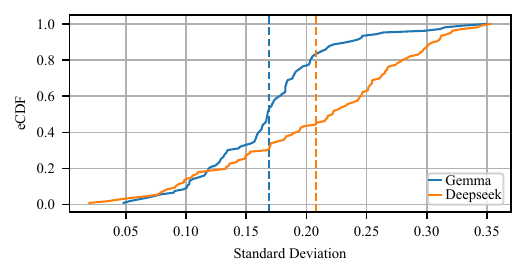}
    \caption{Semantic consistency analysis of answers $A_i$ generated by \ac{LLM} using Gemma and DeepSeek.}
    \label{fig:consistency_sem}
\end{figure}
\Cref{fig:consistency_sem} compares the standard deviation of cosine similarity scores between consecutive outputs for the same question.
As it can be noted, Gemma shows higher semantic consistency than DeepSeek, indicating more stable and deterministic outputs with less variability in meaning when answering the same prompt multiple times.

\subsubsection{LLM-as-a-Judge}

\begin{figure}[b]
    \centering
    \includegraphics[width=.9\columnwidth]{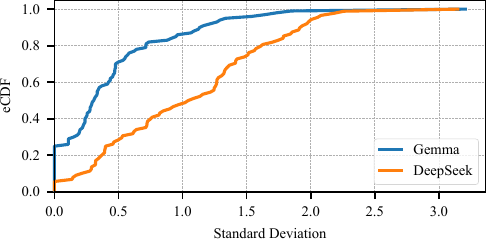}%
    \caption{Comparison of Gemma and DeepSeek model performance, showing the Empirical Cumulative Distribution Function (eCDF) of Standard Deviation.}
    \label{fig:consistency_llm}
\end{figure}

The consistency of the \ac{LLM}-as-a-Judge was evaluated by measuring the standard deviation of the scores it assigned across repeated evaluations of the same question.
\Cref{fig:consistency_llm} shows that Gemma receives more consistent judgments than DeepSeek. Specifically, over \SI{80}{\percent} of Gemma's judgment standard deviations fall below 1.0, whereas DeepSeek's curve rises more slowly, indicating greater variability in its scores.
This suggests that the \ac{LLM}-as-a-Judge is more stable and confident when evaluating Gemma's outputs. A likely explanation is that Gemma's responses maintain a more consistent level of quality and clarity, leading to fewer ambiguities in assessment, while DeepSeek's more variable output quality results in fluctuating judgments.

%
\acresetall

\section{Conclusion}\label{sec:conclusion}

This paper presented a comprehensive evaluation of two state-of-the-art open-source \acp{LLM}~-- Gemma 3 27B and DeepSeek R1 32B~-- on a set of challenging, factual and reasoning-based questions derived from advanced wireless communications.
Our goal was to assess their ability to serve as reliable AI assistants in the telecommunications domain, where correctness and consistency are essential.
By constructing a rigorous benchmark from authoritative textbook material, we were able to assess the quality and the stability of the models' outputs using a mix of lexical metrics, semantic similarity, and subjective scoring through \ac{LLM}-as-a-Judge.

Our results reveal a nuanced performance landscape.
Gemma consistently outperforms DeepSeek in terms of semantic fidelity, lexical precision, and \ac{LLM}-rated accuracy, producing more fluent and technically relevant responses.
However, when it comes to lexical consistency, i.e, the ability to reproduce similar answers across multiple generations, DeepSeek proves to be more stable, with lower variation in BLEU, METEOR, and ROUGE-L scores.
Interestingly, Gemma's outputs receive more consistent evaluations from the \ac{LLM}-as-a-Judge, indicating higher clarity and less ambiguity in how its answers are perceived.
Together, these findings suggest a trade-off between raw answer quality and output determinism, highlighting the importance of evaluating both correctness and consistency for technical deployments.

Future work will focus on expanding the benchmark to include questions related to 5G and 6G standards, thereby broadening the coverage of real-world telecommunications scenarios.
In addition, we plan to evaluate closed-source models, such as GPT-4 and Claude, within the same framework to provide a more complete view of current model capabilities.
Finally, we aim to incorporate human-in-the-loop feedback mechanisms to support iterative refinement and trust calibration, enabling more reliable deployment of \acp{LLM} in high-stakes engineering tasks.
Looking further ahead, we also see potential in evaluating \acp{LLM} within agent-based paradigms, where techniques such as \ac{CoT} and \ac{RAG} could play a central role.


\printbibliography


%



\end{document}